# The simplest rout to generating a train of attosecond pulses

Kazumichi Yoshii, John Kiran Anthony, and Masayuki Katsuragawa*

*Department of Engineering Science, University of Electro-Communications, 1−5−1 Chofugaoka, Chofu-shi, Tokyo, Japan 182−8585*

We report on new routes to generating a train of attosecond pulses. The methods are extremely simple and robust; we need to place only a few thin dispersive materials in the optical path. We numerically demonstrate the generation of a train of attosecond pulses with a transform-limited pulse duration of 728 as and a repetition period of 8.03 fs in gaseous parahydrogen.

As a result of studies of high harmonic generation that have continued extensively since the 1980s and have been boosted by the matured ultrafast technology represented by the Ti:sapphire laser, "attosecond science" is flourishing [1]. Coherent attosecond laser sources have been realized as part of this research, and on the basis of such technology various lines of research centered on the ultrafast dynamics of electrons in atoms or molecules are being developed [1]. In recent years, through the powering of such attosecond laser sources, nonlinear optics in the attosecond regime is also being opened up [1]. These studies have basically been executed in the extreme ultraviolet regime. On the other hand, in closely related studies in the near-infrared–visible–ultraviolet regime, curious new approaches based on four-wave mixing in whispering-gallery-mode microresonators [2] or adiabatic excitation of Raman transitions [3, 4] are being examined extensively to generate broad, discrete, coherent spectra spanning over an octave and then manipulate them [2, 4]. Here, we discuss novel methods of generating a train of attosecond pulses by controlling such highly discrete coherent spectra. The methods are surprisingly simple and very attractive, especially because their flexibility makes them applicable over a wide range, including in high-power lasers.

We show two independent approaches. The first is an extension of the integer temporal Talbot (ITT) concept [5, 6] and can be applied to ultrabroad bandwidths over PHz. The ITT method has essentially been discussed in terms of the generation of a pulse train in the picosecond regime (spectral bandwidth: tens of GHz), mainly regarding application to optical communications technology. The spectral phase of discrete spectral components can be given by Eq. (1).

$$\Theta(\omega_m) \cong \left\{\theta^{(0)} + \theta^{(1)} m\Delta\omega\right\} + \frac{1}{2!}\theta^{(2)}(m\Delta\omega)^2 \quad (1)$$

The point of the idea is that, in dispersion control to form a Fourier-transform-limited (TL) pulse train, each of the spectral phase dispersion, $\theta^{(2)}(m\Delta\omega)^2/2!$, of all discrete spectral components, $\omega_m : \omega_0 + m\Delta\omega$ ($\omega_0$, center frequency; $m$, integer; $\Delta\omega$, frequency space), is controlled so that it is integer multiples of π by adding a positive second-order dispersion (group velocity dispersion) of materials.

The ITT method is very attractive because it is applicable regardless of the sign of the chirp initially included in a pulse, and it does not need the special structured devices providing negative dispersion that are as employed in popular ultrashort-pulse-generation techniques. The ITT method can function well in cases where the spectral discreteness is fairly high (Δω > ~ GHz). For example, in optical communications technology, the spectral spacing is typically located in a range of tens of GHz; a standard kilometer-length optical fiber can therefore become a convenient tool for providing appropriate positive second-order dispersion using the ITT method.

The ITT method can also be applied to spectra with much wider frequency spacings. However, we meet a limitation when the spacing reaches about 10 THz (pulse duration: tens of fs) [7]. In such a situation, the high-order dispersions are not trivial and destroy the ITT condition. This difficulty can be overcome by extending the original ITT concept further so that it includes high-order dispersions.

As a typical example, we discuss a discrete spectrum (bandwidth: 2 π × 496 THz) consisting of five longitudinal modes with an angular frequency spacing, Δω, of 2 π × 124 THz (Fig. 1b). Equation 2 expresses the spectral phase shift, $\Phi(\omega) = n(\omega)\omega x/c$, given through a dispersive material with a refractive index of $n(\omega)$ and a length of $x$. $c$ is the speed of light in vacuum, and $m$ is a longitudinal mode number from −2 to 2.

$$\Phi(\omega_m) = \left\{\phi^{(0)}x + \phi^{(1)}xm\Delta\omega\right\} + \frac{1}{2!}\phi^{(2)}x(m\Delta\omega)^2 \quad (2)$$
$$+ \frac{1}{3!}\phi^{(3)}x(m\Delta\omega)^3 + \frac{1}{4!}\phi^{(4)}x(m\Delta\omega)^4$$

When the spectrum spreads over several hundreds of THz, the spectral phases are substantially influenced by material high-order dispersions (Fig. 1b). Even in such cases, however, as understood from Eq. 2, if the high-order dispersion terms, $\phi^{(h)}x(m\Delta\omega)^h/h!$ ($h \geq 3$), can be simultaneously controlled at integer multiples of π, as is the case with the second-order term, then the spectral phases $\Phi(\omega_m)$ can also satisfy the ITT condition. Although it is difficult to realize this situation in the conventional approach, which employs a single dispersive material, if we introduce different species of dispersive materials corresponding to the term number of the high-order dispersions to be considered, then in principle the ITT condition can be satisfied again.

One of the points of this new approach is that the coefficients, $\phi^{(p)}$, in Eq. 2, are determined by fitting the refractive indices at the respective longitudinal modes, $n(\omega_m)$, with a fourth-order polynomial in each dispersive material. Namely, if we consider spectral phases only at the discrete longitudinal modes, then even if we increase the spectral width we can restrict the dispersion terms to be included to four at most—the mode number minus one, in the case where the mode number is five. The spectral dispersion curve obtained here differs from the original one, still giving us the correct phase values at the longitudinal modes. (This is the extended-ITT method.) Equation 3 represents the relationship that satisfies the ITT condition in the case where the fourth-order dispersion term is the highest and three different species of dispersive material (thicknesses: $x_1$, $x_2$, $x_3$) are employed:

$$\begin{cases} (\phi_1^{(2)}x_1 + \phi_2^{(2)}x_2 + \phi_3^{(2)}x_3)\Delta\omega^2 / 2! = q^{(2)}\pi \\ (\phi_1^{(3)}x_1 + \phi_2^{(3)}x_2 + \phi_3^{(3)}x_3)\Delta\omega^3 / 3! = q^{(3)}\pi \\ (\phi_1^{(4)}x_1 + \phi_2^{(4)}x_2 + \phi_3^{(4)}x_3)\Delta\omega^4 / 4! = q^{(4)}\pi \end{cases} \quad (3)$$

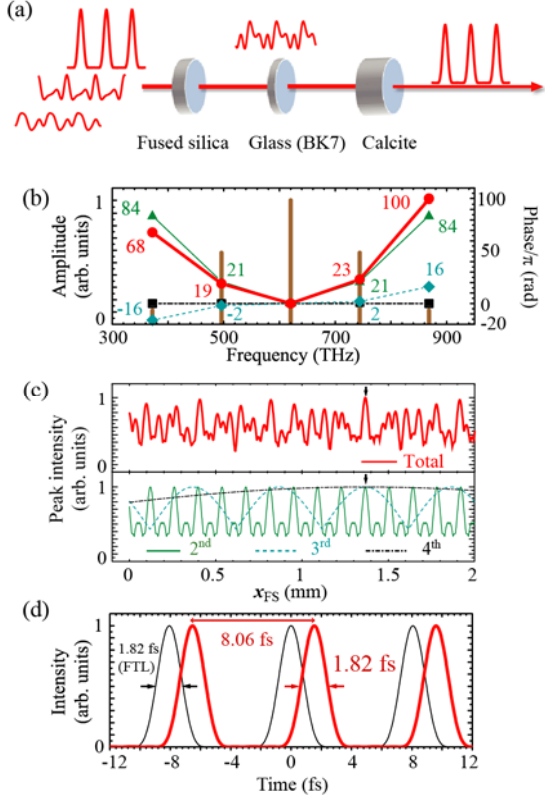
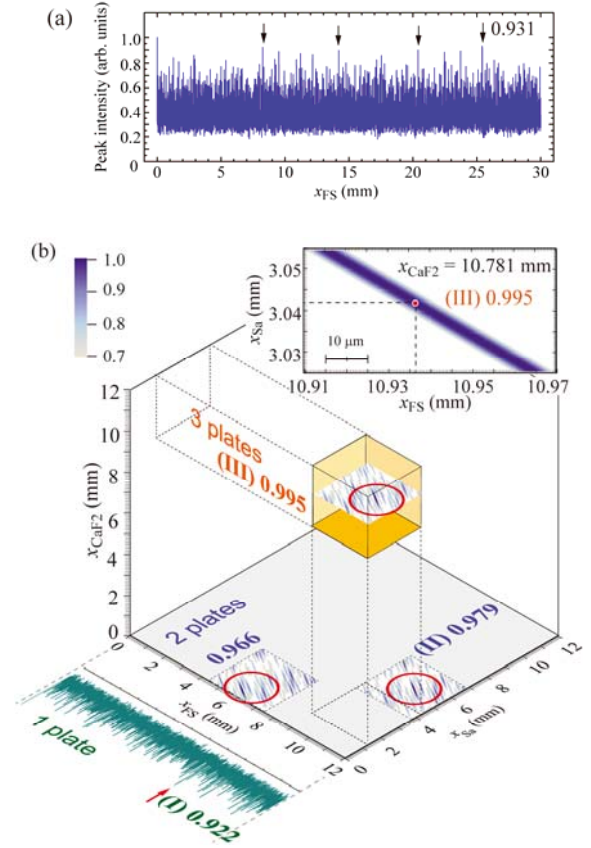

Fig. 1. TL pulse generation with the extended-ITT method. (a) Conceptual illustration of the method. (b) Power spectrum and its spectral phases at the second- (thin green line), third- (dashed blue line), fourth- (dashed and dotted black line) order dispersions, and their total (red line). (c) Peak intensity variation of waveforms as a function of the thickness of the fused silica-glass plate. (d) Temporal intensity waveform realized by applying the extended-ITT method (thick line).

Fig. 2. Exploration of optimum solutions using the NEC method. (a) Peak variation in the intensity waveforms as a function of the thickness of the fused-silica-glass plate. (b) The optimum solutions found on the basis of the NEC method for the three cases of (I) one plate; (II) two plates, and (III) three plates.

The subscripts indicate the species of the dispersive materials employed, and the $q^{(p)}$ values are arbitrary integers.

In practice, we studied the case where three materials—borate glass (BK7), silica glass, and calcite crystal (extraordinary axis)—are employed (Fig. 1a). When we set the integers in Eq. 3 at $q^{(2)} = 21$, $q^{(3)} = 2$, and $q^{(4)} = 0$, we obtain material thicknesses of $x_{FS} = 1.367$ mm, $x_{BK7} = 1.075$ mm, and $x_{Calcite} = 0.179$ mm. Figure 1b displays the spectral phases realized for the second (thin line), third (dashed line), and fourth (dashed and dotted line) orders. The spectral phases at all of the longitudinal modes are controlled to integer multiples of $\pi$ for each of the dispersion-order terms; thereby, the ITT condition is satisfied for the total of the spectral phases, $\Phi(\omega_m)$.

To clarify the precision required for the material lengths, we fixed two of the material lengths at $x_{BK7} = 1.075$ mm and $x_{Calcite} = 0.179$ mm, and we plotted the peak values of the normalized waveform, $I(t) = \left| \sum_m A_m Exp[i\{\omega_m t - \Phi(\omega_m)\}] \right|^2 / I_{TL}$, as a function of the thickness, $x_{FS}$ (Fig. 1c, upper panel). $A_m$ and $I_{TL}$ are the spectral amplitude and the peak intensity to be obtained under the TL condition, respectively. Assuming the practical ITT condition to be max$[I(t)] > 0.99$, then this is satisfied for a range of $x_{FS} = 1.367$ mm $\pm 5.95$ μm. In reality this is a sufficiently controllable range of amount.

To better understand the mechanism of this extended-ITT, we also plotted variations in the peak intensities of the waveforms in the lower panel; each dispersion-order term is included alone. The ITT condition for each dispersion-order term appears with its own periodic cycle. The solution in Eq. 3 is realized when all periodic cycles coincide ($x_{FS}$ = 1.367 mm). Figure 1d is a pulsed waveform obtained under this condition. A TL pulse train with a duration of 1.82 fs and an interval of 8.06 fs is produced (red curve in Fig. 1d; the peak shift is due to the first-order term.)

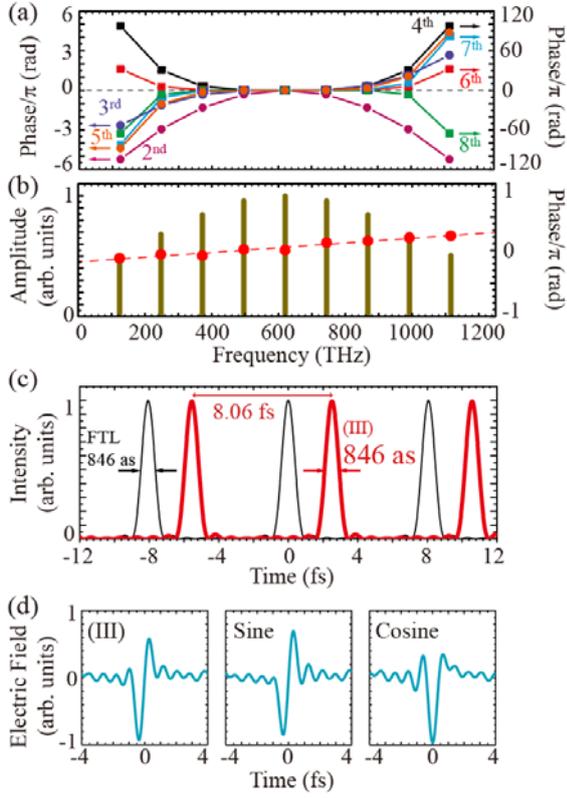
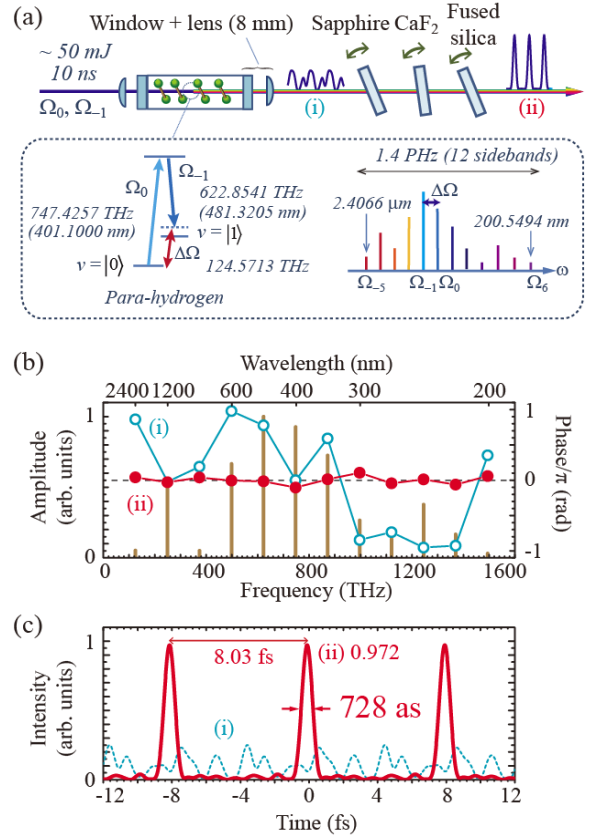

Fig. 3. Attosecond pulse generation by applying the NEC method. (a) Spectral phases in the respective dispersion orders under the optimum condition (III) in Fig. 2b. (b) The total spectral phase and the power spectrum. (c) The temporal intensity waveform (thick line) achieved. (d) Manipulation of the electrical amplitude waveforms around condition (III).

This example shows that the extended-ITT method can be successfully applied to arbitrary initial spectral phases; moreover, the condition satisfying TL conditions completely can be analytically determined. One of the curious findings here is that, as shown in this typical example, the extended-ITT method can function reversely as a simple technique that employs a few thin dispersive materials when it comes to such ultrabroad bandwidth (> several hundred THz) that the high-order dispersion terms are as dominant as the second-order term.

If we increase the longitudinal-mode number further, it is also possible to produce an attosecond pulse train. However, as we increase the mode number we run into increasing difficulty: the species of dispersive materials must be increased correspondingly, and the thicknesses required tend to be unrealistic. Next, we describe the second method, which is more practical and has a potential to give a shorter pulse duration. Technically, like the extended-ITT method (Fig. 1a), this method also simply places one or more dispersive materials in the optical path. Here, however, we abandon to obtain an exact TL condition, and we numerically explore those that approximately satisfy the TL condition by randomly changing the thicknesses of the dispersive materials (Numerical Exploration of pulse Compression: NEC). Surprisingly, we find a lot of such conditions within a realistic range. This method is very similar to that in the extended-ITT, but it essentially differs from it in terms of its physics.

Fig. 4. Numerical experiment for attosecond pulse generation in an actual system (gaseous parahydrogen). (a) Schematic illustration of the setup and the scheme of Raman sideband generation. (b) Power spectrum of the 12 Raman sidebands generated and their phases (i) initially (blue circles) and after applying the NEC method (red dots). (c) The temporal intensity waveforms obtained.

As a typical example, we discuss a discrete spectrum consisting of nine longitudinal modes (bandwidth: $2\pi \times 992$ THz; see Fig. 3b). To understand the basic nature of this method, we first placed only a single dispersive material (silica glass) on the optical axis. We then studied the behavior of the waveform produced as we changed the thickness of the material, $x_{FS}$, continuously. Unexpectedly, even though we introduced only a single dispersive material, peak intensities were recovered at more than 90% of that under the TL condition and appeared four times in a thickness range of 30 mm (Fig. 2a; arrows: 93.1% at 25.461 mm).

If we further increase the number of species used as dispersive materials, the control freedom increases correspondingly; thereby we can naturally expect to find conditions nearer to the TL. We subjected three dispersive materials—silica glass, sapphire crystal (ordinary axis), and calcium fluoride crystal—to the above-described process and explored the optimum conditions by randomly changing the thicknesses of the materials (Fig. 2b). In this exploration, we used 'the random search method' for numerical optimization.

As the target function, we used the peak value of the normalized waveform produced. Here, $\Phi(\omega_m) = \{x_{FS} n_{FS}(\omega_m) + x_{Sa} n_{Sa}(\omega_m) + x_{CaF2} n_{CaF2}(\omega_m)\}\omega_m/c$. In the cubic volume of (12 mm)$^3$ set for the exploration, we found many thickness parameters that gave peak intensities of more than 90% of that under the TL condition. We mapped the points or regions, or both, that had especially high values (Fig. 2b). (The inset depicts an optimum region around condition III, appearing as an oblong shape of 3.18 × 883 µm$^2$). As expected, the peak intensities approached the TL condition (unity) more closely as we introduced more species as dispersive materials. With I (one material), we obtained 0.922 for $x_{FS}$ = 8.254 mm; with II (two materials), we obtained 0.979 for $x_{FS}$ = 10.652 mm and $x_{Sa}$ = 5.312 mm; and with III (three materials), we obtained 0.995 for $x_{FS}$ = 10.937 mm, $x_{Sa}$ = 3.042 mm, and $x_{CaF2}$ = 10.781 mm.

If we had applied the first approach (the extended-ITT method) to the above case, seven species of dispersive material would have been necessary. The point of the second approach is that if we abandon to obtain an exact TL condition, by using a more practical technique of simply adjusting a few thin dispersive materials, we could realize a pulse train that nearly satisfied the TL condition, even for many spectral modes with an ultrabroad bandwidth. In addition, this second approach to the optimum conditions can be established routinely as a numerical exploration method.

Figures 3a and 3b show that this second method essentially differs from the extended-ITT method in terms of its physical nature. Neither of the high-order dispersions under optimum condition III in Fig. 2b satisfies the ITT condition. Nonetheless, their total almost realizes a linear relationship, so their deviations from the ITT conditions cancel each other out. The situation that the bandwidth is ultrabroad, results in substantial inclusion of many dispersion-order terms, which can then reversely function to cancel each other out. Figure 3c shows the pulsed waveform produced. An attosecond pulse train with a pulse duration equivalent to that under the TL condition (846 as) was produced on the basis of this numerical exploration of optimum conditions. This method is very robust. It yields essentially the same result as if we had assumed arbitrary spectral phases initially or employed other dispersive materials.

When we treat a discrete spectrum of which the carrier-envelope offset frequency is controlled [8, 9], we can simultaneously manipulate the carrier-envelope phase in addition to the intensity waveform [10, 11]. Figure 3d demonstrates that the electric field is manipulated to sine- (−1.0 µm) or cosine-like (+1.6 µm) monocycle waveforms by slightly shifting the thickness of the silica glass employed in the NEC method from condition III.

Lastly, we demonstrate the generation of an attosecond pulse train by executing a numerical experiment in an actual system. We generated an ultrabroad coherent discrete spectrum (frequency spacing: 124.5713 THz; bandwidth: 2π × 1.4 PHz for 12 modes, 200.5494 nm to 2.406633 µm) by adiabatically driving the pure-vibrational Raman transition (see the scheme in Fig. 4a; $v$ = 1 ← 0: 124.5710 THz) in gaseous parahydrogen (5.38 × 10$^{19}$ cm$^{-3}$; interaction length: 5 cm). The excitation intensity was 5.0 GW/cm$^2$, and the two driving-laser fields, $\Omega_{-1}$ and $\Omega_0$, were set to 622.8541 THz (481.3205 nm) and 747.4257 THz (401.1000 nm), respectively; their difference was detuned by −300 MHz from the Raman resonance. This numerical experiment was performed by operating a code based on the Maxwell-Bloch equation in the far-off resonant $\Lambda$ scheme [3, 12]. Figure 4b shows a generated discrete spectrum consisting of 12 modes and their spectral phases (blue circles; after the 5-mm-thick silica window and the 3-mm-thick collimating lens). Next we tried to produce an attosecond pulse train by controlling this ultrabroad discrete spectrum. For spectral phase control we adopted the NEC method. As for dispersive materials we employed three species—silica glass, sapphire crystal (ordinary axis), and calcium fluoride crystal—all of which are sufficiently transparent down to the ultraviolet region. Their thicknesses were set to $x_{FS}$ = 1.747 mm, $x_{Sa}$ = 1.556 mm, and $x_{CaF2}$ = 3.727 mm, respectively; these values were obtained as optimal parameters through the numerical exploration. We show the spectral phase (Fig. 4b, dots) and the intensity waveform (Fig. 4c, thick red line) realized under these conditions. An attosecond pulse train with a TL pulse duration of 728 as (peak intensity: 97.2% of that under the TL condition) was recovered from the initial noise-burst waveform (Fig. 4c, dotted line). We therefore demonstrated that this method could function in an actual, practical, system.

In summary, we discovered two novel ways of producing a train of attosecond pulses. One extends the ITT concept to include high-order dispersions and generates a pulse train completely satisfying TL conditions. The other numerically explores the optimum conditions under which we obtain a pulse train approximately satisfying TL conditions; this method is simpler and more practical than the first one. Furthermore, we performed numerical experiment in a real gaseous parahydrogen system, where a train of attosecond pulses was generated with a TL pulse duration of 728 as through adiabatic driving of the fundamental vibrational Raman transition. The attosecond pulse-generation methods described here are extremely simple, requiring the placement of only a few thin dispersive materials and adjustment of their thicknesses. The methods can be applied robustly to a wide variety of systems and are also attractive in their potential to be adapted for use with high-power lasers.

We acknowledge Y. Ohfuti, Y. Kaneda, and S. Nakamichi for their support on in the numerical calculations. We also thank K. R. Pandiri for valuable discussions. M. K. and K. Y. acknowledge supports by Grant-in-Aid for Scientific Research (A) and Grant-in-Aid for Young Scientists (B), respectively. J. K. A. acknowledges the JSPS for granting postdoctoral fellowship.

*katsura@pc.uec.ac.jp